\documentclass[preprint,12pt]{aastex63}

\newcommand{\kms}{{\rm~km~s$^{-1}$}}

\newcommand{\twco}{$^{12}$CO\ }

\newcommand{\thco}{$^{13}$CO\ }

\newcommand{\sdu}{M$_\odot$ pc$^{-2}$}
\newcommand{\sdusp}{M$_\odot$ pc$^{-2}$\ }

\accepted{April, 14, 2025 by The Astrophysical Journal}
\shorttitle{GMC Pressure Support }
\shortauthors{Lada et al. }
\begin{document}

\title{The Role of Pressure in the Structure and Stability of GMCs in the Andromeda Galaxy}

\correspondingauthor{Charles Lada}
\email{clada@cfa.harvard.edu}

\author[0000-0002-0786-7307]{Charles J. Lada }
\affiliation{Center for Astrophysics | Harvard \& Smithsonian, 60 Garden St, MS 72, Cambridge, MA 02138, USA}

\author{Jan Forbrich}
\affiliation{Centre for Astrophysics Research, University of Hertfordshire, College Lane, Hatfield AL10 9AB, UK}

\author{Mark R. Krumholz}
\affiliation{Research School of Astronomy and Astrophysics, Australian National University, Canberra, ACT2601, Australia}

\author{Eric Keto}
\affiliation{Center for Astrophysics | Harvard \& Smithsonian, 60 Garden St, MS 72, Cambridge, MA 02138, USA}

\begin{abstract}
We revisit the role of pressure in the structure, stability and confinement of Giant Molecular Clouds (GMCs) in light of recently published observations and analysis of the GMCs in the Andromeda galaxy (M31). That analysis showed, that in the absence of any external pressure, most GMCs (57\% by number) in M31 would be gravitationally unbound. Here, after  a more detailed examination of the global measurements of surface densities and velocity dispersions, we find that GMCs in M31, when they can be  traced to  near their outermost molecular boundaries, 
require external pressures for confinement that are consistent with estimates for the mid-plane pressure of this galaxy.
We introduce and apply a novel methodology to measure the radial profile of internal pressure within any GMC that is spatially resolved by the CO observations.  We show that for the best resolved examples in M31 the internal pressures increase steeply with surface density in a power-law fashion with $p_{int} \sim \Sigma^2$. 
At high surface densities many of these extragalactic GMC  profiles break from the single power-law and exhibit upward curvature. 
Both these characteristics of the variation of internal pressure with surface density are in agreement with theoretical expectations for hydrostatic equilibrium at each radial surface of a GMC, including the outermost boundary.  
 
 \end{abstract}


\section{Introduction} 

In a recent study Lada et al. (2024) described resolved \twco and \thco observations of a large population of giant molecular clouds (GMCs)  in the Andromeda galaxy (M 31) that were obtained in a deep, interferometric survey of CO line and dust continuum emission using the Submillimeter Array (SMA).  They derived masses, sizes and velocity dispersions for 162 individual \twco GMCs and 84 \thco ''clumps'' resolved in the SMA survey. 
A surprisingly large fraction (117/162) of the GMCs were found to be characterized by simple, single component \twco line profiles, and thus free of confusion by unrelated clouds along the line-of-sight. 
A simple, straightforward analysis of this subsample indicated that 57\% of these GMCs appeared  gravitationally unbound,  defined as having a ratio of kinetic to gravitational binding energies $>$ 1 (i.e., KE/GE $>$1). This result is similar to a recent finding for GMCs in the Milky Way  (Evans et al. 2021) which suggested that as many as $\sim$ 80\% of GMCs in the Milky Way could be unbound. Such large fractions of unbound GMCs would have significant implications for both the formation and evolution of GMCs and for star formation in both these galaxies. 

However, even if the ratio of kinetic to gravitational binding energy of a GMC exceeds 1,
it remains possible that the GMC could be still 
confined by external pressure. Moreover, the inclusion of an external (surface) pressure term in a standard virial analysis might even indicate that some or all of the gravitationally unbound GMCs in M31 are in a state of pressurized virial equilibrium (e.g, Field, Blackman \& Keto 2011). The concept of pressure confinement for molecular clouds has a long history (e.g., Keto \& Myers 1986; Maloney 1988, 1990; Elmegreen 1989; Betoldi \& McKee 1992) but until recently has been lacking compelling empirical support. However, a recent study by Keto (2024) has provided  intriguing evidence that the inner \thco emitting regions of Milky Way GMCs are on average in approximate virial equilibrium with the pressure of surrounding gas across any choice of cloud boundary. If extrapolated  to outer cloud regions this result would be  consistent with the notion that Galactic GMCs may be entirely confined by the mid-plane pressure of the Milky Way.
This would contradict recent suggestions, based on a virial analysis that ignores external pressure,  that most Milky Way GMCs are unbound (Evans et al., 2021).

Incorporating empirical knowledge of the internal pressure structures of GMCs alongside the traditional metrics of cloud kinetic  and gravitational binding energies significantly enhances our ability to critically assess their dynamical states. This, in turn, allows for a deeper understanding of the fundamental nature of these complex and important astrophysical objects.

In this paper, we introduce a new methodology for measuring the 
spatial pressure profiles within individual, spatially resolved GMCs and apply it to the GMCs of the Andromeda galaxy (M31). Our analysis shows that the internal structures of the best-resolved clouds in M31 are in broad agreement with theoretical expectations for hydrostatic equilibrium  at each measured cloud radius. Moreover, in cases where we were able to trace the clouds to near their outermost molecular boundaries, we find that the external pressures required to maintain such equilibrium and confine the GMCs are consistent with the mid-plane pressure derived for M31.

This paper is the first in a series of three exploring the dynamical properties of the Giant Molecular Clouds (GMCs) in Andromeda. The second and third papers focus on the more detailed theoretical implications of the findings in this paper. Specifically, the second paper will examine how these results help distinguish between global collapse and globally supported models of GMCs (Krumholz et al. 2025), while the third will test how well hydrostatic equilibrium models match observational data for individual GMCs in Andromeda (Keto et al. 2025).

\section{Observations}

\subsection{The SMA CO Survey of GMCs in Andromeda}

The observations analyzed in this paper were obtained as part of an SMA large program dedicated to a simultaneous survey of 230 GHz continuum and CO line emission from individual molecular clouds across the Andromeda galaxy. A basic description of the receiver and spectrometer configurations and calibration has been presented in the three earlier papers describing the survey results  (Forbrich et al. 2020, Viaene et al. 2021, Lada et al. 2024). Briefly, the SMA survey of Andromeda consisted of 80 individual SMA pointings  acquired over four consecutive fall seasons (2019-2022). At the observing frequency the six-meter dishes of the array provided a primary beam (FWHM) of 55 arc sec or $\sim$ 200 pc. The interferometric spatial filtering provided an approximate maximum recoverable scale of $\sim$ 100 pc. The observations were obtained with the eight-antenna array in the subcompact configuration providing a synthesized beam size of $4.5'' \times 3.8''$ at 230 GHz corresponding to a spatial resolution of $\sim 17 \times 14$~pc in M31. The survey was designed to obtain deep and uniform sensitivity in the continuum band images for all observed fields. The noise in the CO moment-zero (integrated intensity) images, used for analysis, tended to be somewhat higher and less uniform than in the continuum images due to confusion from  varying amounts of extraneous CO emission in the fields. This, coupled with variations in weather and in instrumental capabilities (e.g., the occasional absence of one or two antennas that were not in service due to maintenance and/or technical issues) resulted in somewhat degraded image noise in roughly  30\% of the CO images in the sample. The GMCs identified by the SMA observations were defined as objects having \twco emission in excess of the 3-$\sigma$ image noise within a set of contiguous pixels whose area was at least partially resolved by the synthesized beam. 

\section{Results and  Analysis: The Role of Gravity and Pressure in Andromeda's  GMCs }

\subsection{Pressure Confined GMCs?}

\begin{figure}[t!]
\centering
\includegraphics[width=.8\hsize]{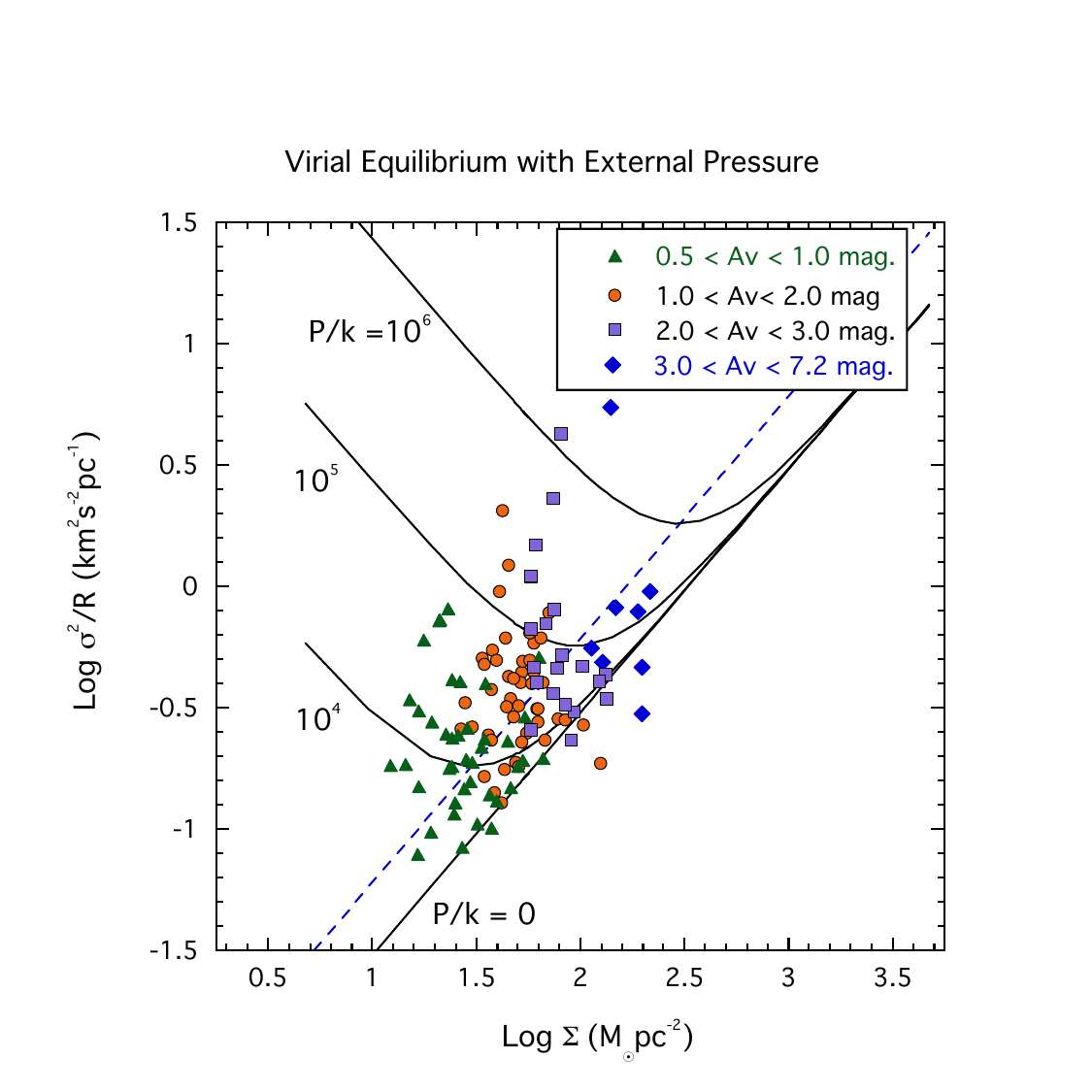}
\caption{The pressurized viral equilibrium diagram for GMCs in M 31. 
The solid lines are the theoretically predicted relations for GMCs in virial equilibrium with external surface pressure for different varies of the external pressure (i.e.,  p/k = 0, 10$^4$, 10$^5$, and 10$^6$). The dashed line divides unbound and bound clouds in the absence of any external pressure (i.e., p/k = 0). The 
 data color-coded by the value of the outer boundary surface density, $\Sigma_0$, used to define each cloud. The required pressure for confinement spans a range of approximately two orders of magnitude and appears to be positively correlated with the cloud's  outer boundary surface density which is set by the image sensitivity.  See text for discussion.  \label{KetoDiagramM31}  }
\end{figure}

To investigate the role of pressure in confining GMCs we first analyze the virial diagram (Keto \& Myers 1986) for the M 31 GMC population, shown here in Figure \ref{KetoDiagramM31}. This diagram plots the quantity $\sigma^2/R$ vs $\Sigma_{GMC}$ for the surveyed GMCs.  Here $\sigma$ is the measured velocity dispersion of the molecular gas and $R$ is a measure of the GMC size, that is, the spatial scale encompassing all the gas, expressed as a radius (i.e., $R = \sqrt{A/ \pi }$, where $A$ is the area of the GMC).  The quantity $\sigma^2/R$ has the units of acceleration and, assuming that the velocity dispersion is due to turbulence, is equivalent to a measure of the pressure per unit surface density (i.e., $p/\Sigma$). It is a useful diagnostic to describe the virial equilibrium state of a whole molecular cloud in the presence of an external pressure term.  Also plotted in the figure are curves (solid hyperbolic lines) depicting the theoretically predicted loci of GMCs in virial equilibrium with an external pressure for different values of external pressure (i.e., p$_{ext}$/k = 0, 10$^4$, 10$^5$, and 10$^6$). A dashed line marks the 
boundary between gravitationally bound and unbound clouds where the kinetic energy equals the gravitational binding energy (i.e., KE/GE $=$ 1) and a straight solid line marks the location of virial equilibrium (i.e., 2KE/GE $=$ 1) in the absence of any external pressure. To place these lines on the diagram we calculated the gravitational binding energy assuming a spherical cloud model with a density gradient $\rho(r) \sim r^{-1}$ following Lada et al (2024). As can be seen here a significant fraction of the GMCs appear to be unbound in the absence of an external pressure corroborating the findings of Lada et al. (2024). \footnote{However, the exact fraction of sources that appear unbound (or bound) is very uncertain since the exact locations on the plot of the boundary between bound and unbound clouds and the line of virial  equilibrium are sensitive to the assumed geometry and density structure of the adopted cloud model while the precise locations of the data are sensitive to systematic uncertainties in the CO mass calibration.} 
 
External pressures ranging from p/k $\sim$ 10$^4$ to 10$^6$, would be required for all of the unbound clouds in figure \ref{KetoDiagramM31} to be in virial equilibrium with, and confined by, an external surface pressure. The most likely source of such external pressure for these GMCs would be the pressure in the mid-plane of the galaxy's disk produced by the weight (including stars and gas) of the disk itself (Elmegreen 1989; Blitz \& Rosolowsky 2004, Keto 2024). However, the range of external pressure required for confinement exceeds the expected variation of the mid-plane pressure in a disk galaxy like M 31. This pressure  approximately varies with the square of the total gas (HI $+$ H$_2$) surface density, $\sim {\pi\over{2}} G \Sigma^2$  (Krumholz \& McKee 2005).
  Between 6 and 16 kpc, where 90\% of the GMCs in the SMA survey of M 31 are found, the total gas surface density (HI $+$ H$_2$) only varies by about a factor of $\sim$ 2 (Johnson et al. 2016) corresponding to an expected  mid-plane pressure variation of about a factor of $\sim$ 4. This is well short of the spread in the data in the figure which would require pressure variations of more than two orders of magnitude to virialize all the clouds across the observed range of galactocentric distances. 
 
We can directly estimate the mid-plane pressure in the disk of M31 from observations using the following equation (Elmegreen 1989): 
\begin{equation}
 p_{mp} = p_{ext} \approx {\pi\over{2}}G\Sigma_g(\Sigma_g + \Sigma_* {\sigma_g\over{\sigma_*}}).
\end{equation}
\noindent
Here $\Sigma_*$ is the stellar surface density in the disk and $\Sigma_g$ is the total (HI + H$_2$) gas surface density; $\sigma_*$ and $\sigma_g$ are the velocity dispersions of the stars and gas, respectively. Because M 31 has been extensively studied all the quantities in this equation can be obtained from the literature. We evaluate equation 1 using values for a galactic radius of 10 kpc.
We adopt $\Sigma_g = $ 10 \sdu, $\sigma_g =$ 9 \kms (Johnson et al. 2016)  and $\Sigma_* =$ 210 \sdu\  (Tamm et al. 2012, Rahmani et al. 2016). 
The stellar velocity dispersion in the disk is found to be an increasing function of stellar age with $\sigma_* = $ 30 \kms\   for main sequence stars and $\sigma_* = $ 90 \kms \ for red giant stars (Dormann et al. 2015). Here we adopt $\sigma_* = $ 90 \kms\  to obtain a lower limit to the mid-plane pressure. From equation 1 we find $p_{mp} \gtrsim $ 1.0 $\times$ 10$^4$ $k_b$. Here $k_b$ is the Boltzmann constant. 

Evidently, the mid-plane pressure of the M 31 disk cannot virialize or confine all the unbound clouds observed across M 31. However, some care is warranted before taking the derived CO surface densities at face value. In particular, earlier studies (Lada \& Dame 2020, Lewis et al 2021) have demonstrated that the derived surface densities of GMCs are quite sensitive to a systematic in the methodology used to define the clouds. The surface density of a GMC is defined by the ratio of its molecular mass to its area. Measurements of both quantities in a GMC depend on the outer surface density boundary ($\Sigma_0$) empirically adopted for the cloud. For the M 31 data used here these boundaries are set by the (3-$\sigma$) level of the image noise of the corresponding data and, although designed to be very similar, the image noise levels were found to vary due to the various factors discussed earlier. In Figure \ref{KetoDiagramM31} the data are color coded to indicate the various (noise fixed) values of $\Sigma_0$ used to define the clouds in the  M 31 sample. 
 
\begin{figure}[t!]
\centering
\includegraphics[width=.8\hsize]{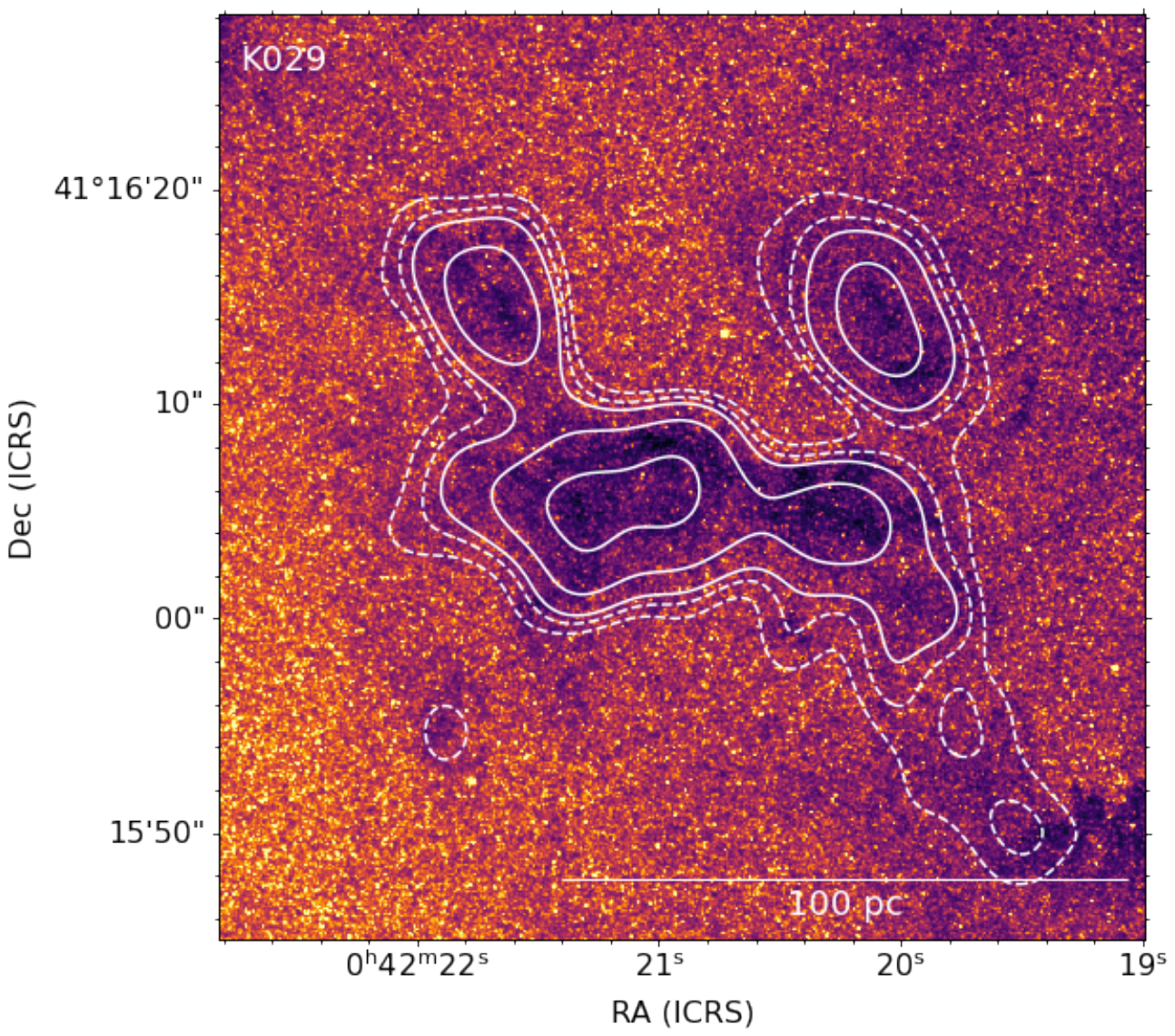}
\caption{The surface density map derived from the J=2-1 \twco observations of the GMCs K029A and B superimposed on an optical (F475W) HST image of the corresponding region illustrating the close correspondence between the extents of the observed CO emission and extinction produced by the clouds. The lowest two dashed contours correspond to iso-density contours at the 1 and 2 $\sigma_{noise}$ levels of image noise. The lowest solid contour corresponds to the 3 $\sigma_{noise}$ defining boundary for the clouds. The surface density contours correspond to values of 0.93, 1.9, 2.8, 5.6 and 11.2 visual magnitudes or 18, 36, 54, 108, and 216 \sdu, respectively.  In this field the 1 $\sigma_{noise}$ level is very close to the expected physical (molecular) boundary (A$_{\rm v} \approx $ 0.5 magnitudes) of the two clouds.   See text for discussion.  HST image from Dalcanton et al. (2012).\label{overlayK029}  }
\end{figure}

 The figure shows that the location of GMCs on the diagram systematically varies with the adopted outer boundary, $\Sigma_0$. The GMCs with the lowest boundary surface density (expressed here in terms of equivalent magnitudes of visual extinction, A$_{\rm v}$), appear to require the lowest external confining pressures. The adopted surface density boundaries of these particular clouds are equal to or less than A$_{\rm v}$ = 1 magnitude (i.e., $ \lesssim$  22 \sdu) and these boundaries are likely very close to the expected physical boundaries of the GMCs.\footnote{The edge of a molecular cloud is defined by the location of the atomic-molecular phase transition in the gas. In the Milky Way this occurs at a surface density of  $\approx$ 0.5 magnitudes (or $\Sigma_{\rm HI-H_2}$ $\approx$ 10 \sdu; e.g., Sternberg and Dalgarno 1995).}   As an illustration of this we present in Figure \ref{overlayK029} the surface density map of the GMCs K029A and K029B superimposed on an HST image of the region. This is one of several areas in M31 where our SMA observations coincided with high-resolution HST images in which cloud-like features of visual extinction were present. The lowest surface density CO contour in the image corresponds to the  1$\sigma_{noise}$  image noise level and is very close to the expected molecular/atomic transition boundary which we consider the physical boundary for  molecular clouds. It is also at the depth of the 3$\sigma_{noise}$ cloud boundaries measured for many clouds in the deepest images of our survey. The figure demonstrates how the CO emission can closely trace the borders of the extinction, even at the image depth of the K029 field.

For the set of GMCs with the deepest measured outer 3$\sigma_{noise}$ boundaries (A$_{\rm{v}}$ $\leq$ 1 magnitude)
 the pressures required for their confinement,  5 x 10$^3$ $\lesssim$ p/k $\lesssim$ 5 x 10$^4$, are comparable  to the mid-plane pressure we estimated for M31  (p/k $\gtrsim$ 1 x 10$^4$) and to the expected mid-plane pressure (p/k $\approx$ 4 x 10$^4$) of a typical disk galaxy (e.g., Blitz \& Rosolowsky 2006; Leroy et al. 2008, Kasparova \& Zasov 2008). These facts indicate that the mid-plane pressure of M31 is likely sufficient to confine these particular clouds. Moreover, because the depths of these particular measurements are close to the actual molecular boundaries of the GMCs, we have not accounted for any extra pressure exerted by their surrounding atomic envelopes. While the sizes and masses of these GMC envelopes are presently unknown, any pressure exerted by their weight will  reduce the external pressure required for overall confinement  (Elmegreen 1989). This further reinforces the idea that the mid-plane pressure in M31 is very likely adequate to confine these GMCs. 
  
Figure \ref{KetoDiagramM31} also indicates that GMCs with  $\Sigma_0$  $>$ 1 magnitude would require an additional source of external surface pressure, beyond the pressures exerted by the galaxy's mid-plane, and the GMC's surrounding atomic envelope in order to be in (virialized) pressure equilibrium. The higher values of $\Sigma_0$ for these GMCs indicate that the corresponding CO observations are not able to detect their full molecular masses and extents due to elevated image noise in the corresponding fields. That the required external pressures for confinement of these GMCs increase systematically with $\Sigma_0$ suggests that the extra source of required pressure is the weight of the undetected molecular gas outside the adopted (noise-defined) boundaries of these clouds.  

It is therefore plausible that most of the GMCs in M31 are  bound across all internal boundaries, including their outer edges, by a combination of surface pressure and gravity, and that they are in a state of  pressurized virial equilibrium throughout their structure. To test this hypothesis, it would be crucial to determine the radial variation of internal pressure within individual GMCs. Such measurements could provide key insights into whether the GMCs maintain virial equilibrium from their innermost regions to their outer boundaries. In the following section, we will employ new methodology that enables us to measure the internal pressure structure of the best-resolved GMCs from the M31 SMA survey and better evaluate the dynamical natures of these extragalactic molecular clouds. 
  
\subsection{ Measuring the Internal Pressure Profiles within Individual GMCs}

\subsubsection{Methodology}
The GMCs in the SMA survey of Andromeda are all spatially resolved to varying degrees.   From the observations we can then directly measure the internal pressure, $p_{int}$, of a GMC as a function of cloud depth, with the assumption that the velocity dispersion ($\sigma$) is due to turbulent broadening: 
\begin{equation}
p_{int}(r) = \Sigma(r)  {\sigma^2\over{r}}.   
\end{equation}
\noindent
 Here $\Sigma(r)$ is the mass surface density and $r$ the projected radial dimension as defined below. 
Important insight into the role of pressure in the stability and structure of the M31 GMCs can be obtained by analysis of the dependence of internal pressure on cloud depth. To do this we first employ a methodology similar to that we have used to measure the basic global properties of the GMCs. For each GMC we define a sequence of increasing boundary surface densities, $\Sigma_i$, that begins with $\Sigma_i$=$\Sigma_0$, the outermost cloud boundary (set to the nominal 3$\sigma_{noise}$ level of the moment 0 image containing the GMC) and increases by  increments of  $\sigma_{noise}$, until  $\Sigma_i$ encloses an area comparable to that of the synthesized beam of the interferometer (i.e., $r_i \approx R_{beam} \approx {\rm7}$ pc). We then measure the average surface density, $<\Sigma_i>$, and from the average CO line profile, the velocity dispersion, $\sigma_i$, for area corresponding to $\Sigma > \Sigma_i$. Each value of  $\Sigma_i$  corresponds to a radius $r_i$ that is defined by the area ( $r_i = (A_i/\pi)^{0.5}$) of all the pixels with $\Sigma > \Sigma_i$. 
This areal approach for defining the average $ r, \sigma, {\rm and}\ \Sigma$ has the benefit that the derived quantities are independent of cloud geometry. This is particularly advantageous for measurements of GMCs which are generally non-circular and irregularly shaped. 
 Table \ref{12coextraction} displays the resolved basic properties of the individual GMCs in our M31 sample that are defined and derived in this way.
With the surface densities, velocity dispersions and radii in the table we can then use equation 1 to construct an internal radial profile of average pressure for each individual GMC. 

\begin{deluxetable*}{cccccccc}
\tablenum{1}
\tablecaption{Resolved Physical Properties: M31 GMCs\label{12coextraction}}
\tablewidth{0pt}
\tablehead{
\colhead{GMC ID}&\colhead{Av$_i$}&\colhead{Mass$_i$}&\colhead{$r_i$}&\colhead{$\sigma_i$(12)}&\colhead{$\sigma_i$(13)}&\colhead{S/N}&\colhead{$\Sigma_i$}\\
\colhead{ }&\colhead{mag}&\colhead{M$_\odot$}&\colhead{pc}&\colhead{km\ s$^{-1}$}&\colhead{km\ s$^{-1}$}&\colhead{($^{13}$CO) }&\colhead{M$_\odot$ pc$^{-2}$}}
\startdata
K001A&2.7&5.93E+05&37.8&4.04&3.67&15.6&132.44\\
K001A&3.6&5.61E+05&35.0&4.02&3.70&16.6&146.15\\
K001A&4.5&5.29E+05&32.6&4.00&3.72&17.5&158.37\\
K001A&5.4&4.97E+05&30.5&3.97&3.73&18.2&169.61\\
K001A&6.3&4.63E+05&28.6&3.95&3.75&18.6&180.62\\
K001A&7.2&4.23E+05&26.4&3.92&3.77&18.6&192.81\\
K001A&8.1&3.85E+05&24.5&3.87&3.76&18.6&204.54\\
K001A&9.0&3.48E+05&22.7&3.80&3.71&18.6&216.13\\
K001A&9.9&3.12E+05&20.9&3.71&3.62&18.5&227.82\\
K001A&10.8&2.74E+05&19.1&3.60&3.51&18.4&240.41\\
K001A&11.7&2.42E+05&17.5&3.58&3.50&18.0&251.92\\
K001A&12.6&2.11E+05&15.9&3.66&3.57&17.2&263.91\\
K001A&13.5&1.78E+05&14.3&3.79&3.66&16.1&278.22\\
K001A&14.4&1.57E+05&13.2&3.85&3.71&15.7&287.82\\
K001A&15.3&1.37E+05&12.1&3.89&3.75&15.4&297.21\\
K001A&16.2&1.17E+05&11.0&3.92&3.77&15.1&306.93\\
K001A&17.1&9.69E+04&9.86&3.95&3.80&14.8&317.40\\						
\enddata
\tablecomments{ Av$_i$ is the defining surface density boundary for each cloud layer expressed in magnitudes of extinction. R$_i$ is the radius deconvolved from the synthesized SMA beam. S/N is the peak signal-to-noise measured in each $^{13}$CO spectrum. See text for discussion. Only a portion of Table 1 is shown here. The entire table is published in machine readable form in the online version of the paper. }
\end{deluxetable*}

Molecular clouds are typically characterized by what can be described as inwardly increasing surface density gradients, that is, surface densities that increase with decreasing spatial scale or cloud area (e.g., Lada et al. 1994; Alves et al. 1998). As a result,  cloud depth can be usefully measured in terms of either a projected radius or a surface density. For GMCs that are only modestly resolved (i.e., $R_{GMC} \lesssim 10 R_{beam}$), like those in M31, it can be more insightful to measure  GMC depth  in terms of surface density than radius because in  such clouds the dynamic range in surface density is typically much larger than that in radius.

\begin{figure}[t]
  \centering

\includegraphics[width=.7\hsize]{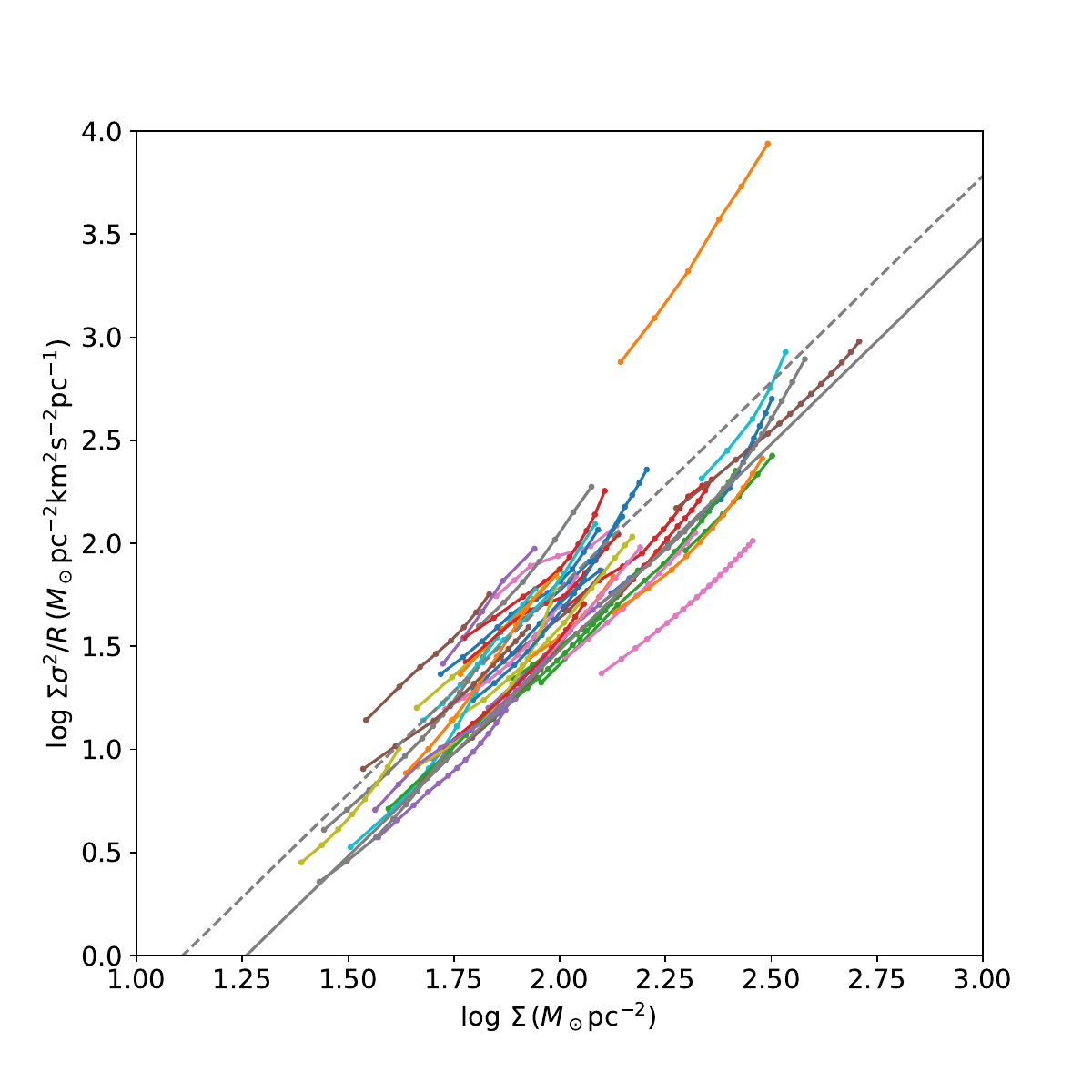}
\caption{Internal pressure verses surface density profiles for the 48 best resolved GMCs in our M31 sample. Each track represents the locus of points for an individual GMC in this space. The individual points in each trace are separated by the 1 $\sigma$ internal uncertainty in each coordinate. Only sources with five or more measurements and single component (unconfused) \twco line profiles are included in this GMC sample. The solid line is the virial relation with no external pressure term. The dashed line marks the boundary between bound and unbound clouds in the absence of a confining external pressure, that is, where the kinetic energy of the gas equals its gravitational binding energy. The GMCs generally exhibit inwardly increasing pressure gradients that are near linear power-laws with $p_{int} = \Sigma \sigma^2/R \sim \Sigma^2_{gas}$.
  \label{PressureDiagramM31_individualGMCs}}

\end{figure}


\subsubsection{Results}

Using our methodology, we constructed the profiles of internal pressure verses surface density for each of the M31 GMCs in our sample. 
Figure \ref{PressureDiagramM31_individualGMCs} shows these individual relations where  the data for each GMC are plotted as a series of points connected by solid traces. The points correspond to different surface density levels within the cloud separated by increasing 1$\sigma_{noise}$ steps in surface density, starting from the value  $\Sigma = \Sigma_0$ for each cloud. Spaced in this way the horizontal and vertical displacements between adjacent points on the plot are equal to the one sigma formal error in each coordinate for each point. However, we should note that with this sampling frequency in surface density, the surface densities and derived pressures are oversampled in radius and consequently we include only clouds with five or more such surface density measurements in the plot. This restricts the sample to a set of 48 GMCs that are both reasonably well resolved and exhibit a significant dynamic range in internal surface density. This represents only $\sim$ 40\% of the GMCs in Figure \ref{KetoDiagramM31}. Because of this requirement, however,  only 11 GMCs appear on the plot that are apparently unbound at all $\Sigma$s, representing only 16\% of the GMCs that were found to be globally unbound by Lada et al. (2024) and in Figure \ref{KetoDiagramM31}. Thirty-seven GMCs appear in the bound region of the plot, accounting for 74\% of the GMCs seen to be globally bound in figure \ref{KetoDiagramM31}.

As a group the GMCs in Figure \ref{PressureDiagramM31_individualGMCs} exhibit reasonably well behaved pressure profiles that correspond to inwardly increasing pressure gradients. These profiles 
are generally parallel to themselves and to  both the virial equilibrium line and the boundary between bound and unbound clouds (i.e., where 2KE $=$ GE, and KE=GE, respectively, and $p_{int} \sim \Sigma^2$). 
The dispersion in the profile locations is relatively narrow. For 33 GMCs with an average surface density within 0.1 dex of log $\Sigma = 2$, the median and standard deviation of internal pressures are found to be $p_{int} =$ log$(\Sigma \sigma^2/R) = 1.69 \pm$ 0.18 dex.  This level of scatter almost entirely can be accounted for by the observed scatter in the dust-calibrated, mass conversion factor ($\alpha(CO) =$ 10 $\pm$ 4.5)  directly measured for the SMA GMC sample (Viaene et al. 2021) and employed to derive the cloud masses used here (Lada et al. 2024).  Thus the intrinsic dispersion of this sequence of pressure profiles is narrower than shown in the figure. Additionally, the median value of these M31 pressure profiles is within 1 $\sigma$ of the virial relation depicted here. 

There is a tendency for many of these pressure profiles to bend upward at the highest surface densities. 
A particularly good example of such a pressure gradient structure is shown in Figure {\ref {PvsSDK001} where we plot $p_{int}$, against $\Sigma$ for the \twco observations of the GMC K001A, a gravitationally bound GMC. The \twco  relation displays a well-behaved, inwardly increasing pressure gradient. 
 At the lower surface densities (i.e., $\lesssim$ 220 \sdu) the pressure displays a smooth linear (power-law) rise with surface density, with a least squares fit yielding $p_{int} \sim \Sigma^{1.88\pm 0.05}$. For this cloud the observed relation is coincident with the virial equilibrium line within the errors.  This demonstrates that the internal structure of the K001 GMC is configured to produce a pressure gradient that maintains the lower surface density, outer regions of the GMC in a state of virial balance at every radius. This is in close agreement with the theoretical expectations for an object in hydrostatic equilibrium where  inward and outward vertical forces are balanced and $p_{int} \sim G\Sigma^2$, independent of cloud shape (Bertoldi \& McKee 1992).

\begin{figure}[]
\centering

\includegraphics[width=0.6\hsize]{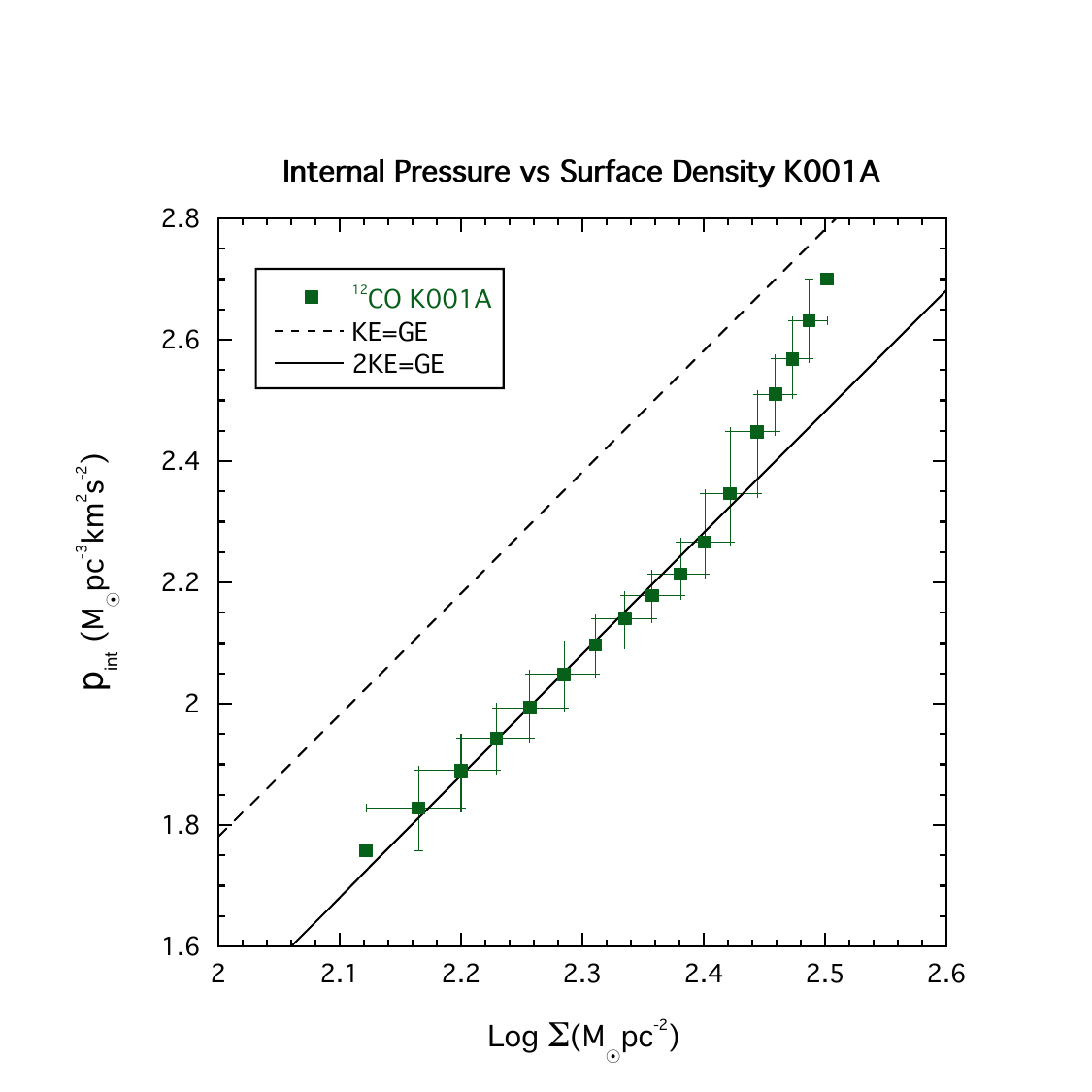}
\caption{Internal pressure--surface density ($p_{int}$--$\Sigma$) relation for the GMC K001. The solid line is the virial relation with no external pressure term. The dashed line is the boundary between bound and unbound clouds in the absence of a confining pressure, that is, where the kinetic energy of the gas (KE) equals its gravitational binding energy (GE). The error bars are the internal errors. The relation is indicative of a state of virial equlibrium across each cloud boundary or radius as would be expected for a cloud in hydrostatic equilibrium. The break away from the simple virial relation at the high surface densities is likely due to the pressure from the weight of the outer cloud layers being more important than gravity for confining the gas in the inner depths of the cloud. See text.
\label{PvsSDK001}} 

\end{figure}

At $\Sigma \approx$ 240 \sdusp the data break from the single linear relation and the pressure increases more rapidly with surface density with $p_{int} \sim \Sigma^{4.4 \pm0.07}$. The higher surface density inner layers of the cloud increasingly depart from the (pressure-free) virial equilibrium relation.  
Indeed, the innermost region of the cloud, with the highest surface density, is situated very close to the boundary between bound and unbound objects. Consequently, the internal pressure is very close to exceeding the limit that would enable the inner cloud material to be bound or confined by gravity alone. It seems highly unlikely that the inner parts of this massive GMC are nearly dynamically unbound, whilst the outer regions remain strongly bound and in virial equilibrium. A more plausible explanation is that the inner regions are pressure-confined by the weight of the cloud’s outer layers.
In  other words, the inner regions, which have lower enclosed mass, are not primarily in virial balance between their internal kinetic and gravitational binding energies. Instead, the balance there is dominated by the pressure exerted by the weight of the surrounding cloud layers. Therefore, the outer and inner regions are likely to be smoothly connected in virial  equilibrium when considering a radially or depth-dependent pressure term.

It is important to consider here that the increase in the internal pressure in the innermost regions is not a result of an increase in the velocity dispersion at the higher surface densities. Examination of our data shows no significant increase in velocity dispersion between the outermost and innermost surface densities in K001A. 
Consequently,  the inward increase in pressure  must be the result of a steeply increasing cloud density gradient. This is reminiscent of the situation for the well studied Galactic dark cloud Barnard 68. B68 is an isothermal cloud whose structure is exceedingly well described by the Lane-Emden equation for hydrostatic equilibrium where the internal pressure gradient is primarily the result of an inwardly increasing density gradient at all cloud radii (Alves et al. 2001, Lada et al. 2003). The main difference here would be that the K001A GMC is in the regime where the internal supporting pressure is turbulent ($\rho$v$^2$) rather than thermal (nkT).

 Finally, we note that the methodology described (3.2.1)  also enables resolved profiles of individual GMCs to be plotted on the virial diagram. A similar analysis (not shown here) using the virial plot closely aligns with the conclusions derived from the pressure diagram.

\begin{figure}[]
\centering
\includegraphics[width=0.5\hsize]{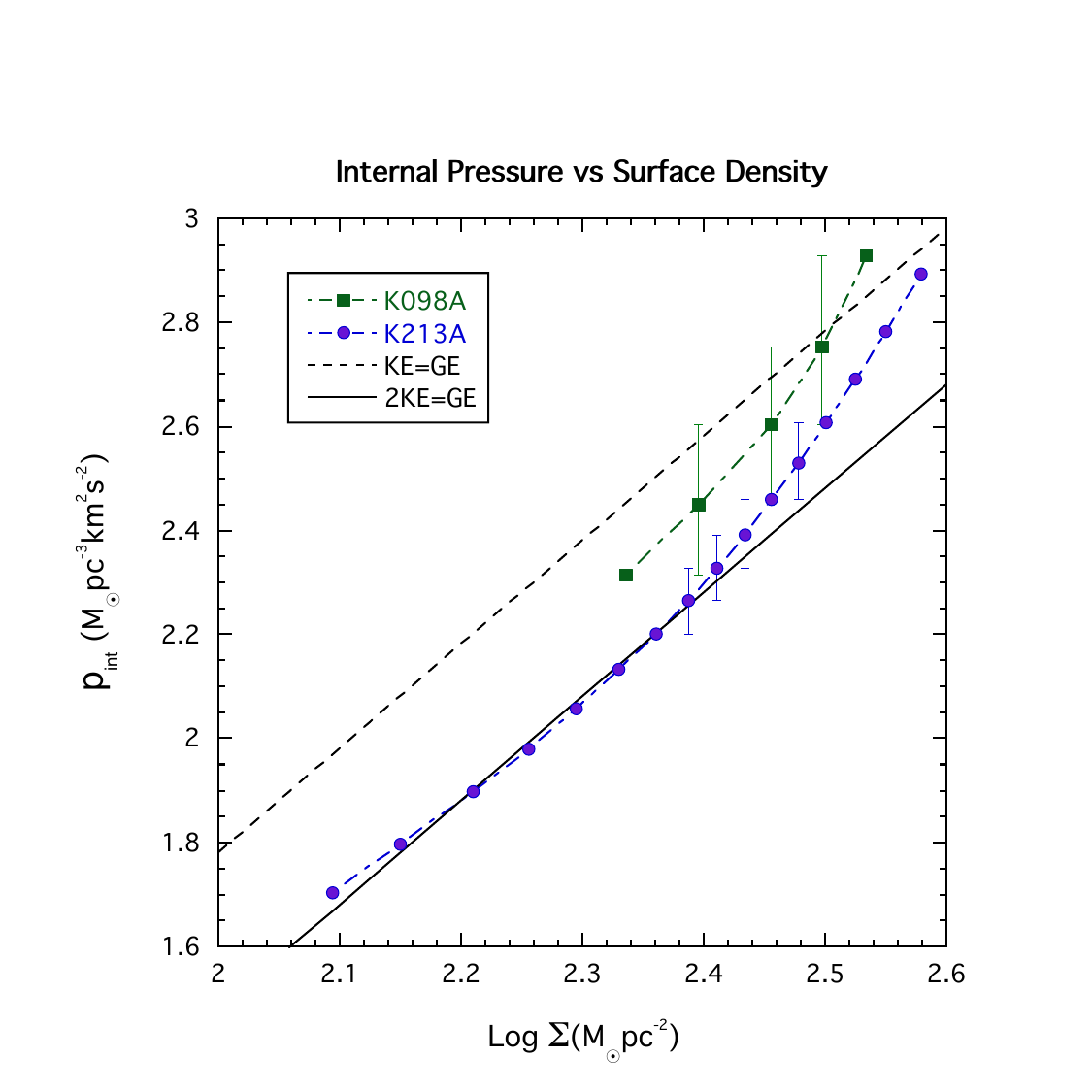}
\caption{Internal pressure--surface density ($p_{int}$-$\Sigma$)  relation for the GMCs K213A and K098A in M31. The shapes of the two profiles are very similar and coincident within the errors, consistent with virial equilibrium at every boundary surface density or radius for each cloud.   Otherwise, same as  figure \ref{PvsSDK001} \label{PvsSD2GMCs}} 
\end{figure}

\section{Discussion}

\subsection{Pressure Confinement and Hydrostatic Equilibrium in the GMC Population}

The analysis presented above provides interesting insights regarding the role of pressure in the stability and support of GMCs in M31.
In particular, GMCs that were traced to  near their outermost boundaries require surface pressures for confinement and virialization that are comparable to or less than the expected mid-plane pressure of M31’s disk. This suggests that even GMCs, whose total kinetic energies appear to exceed their gravitational binding energies, could still be confined by external pressure and be in virial and hydrostatic balance. For GMCs that could not be traced to close to their outermost boundaries, only their inner regions could be measured, and their outer layers remain undetected. In these cases, our analysis found that additional sources of pressure are required to confine and bring the otherwise unbound or loosely bound regions into virial equilibrium. The amount of additional pressure needed scales directly with the adopted boundary for the cloud, leading to the hypothesis that this extra  pressure is provided by the weight of the surrounding, undetected, cloud envelope.

As a  test of this hypothesis we developed a technique to enable us to construct the internal pressure profiles of well resolved GMCs. This enabled a better evaluation of their overall dynamical states. Using this methodology we were able to construct profiles for 48, mostly bound, GMCs. These profiles exhibited a number of very similar properties. They were generally characterized by inward increasing pressure gradients  that are near linear power-laws with  $p_{int}  \sim \Sigma^2$, consistent with the expectations of hydrostatic equilibrium. Many profiles also exhibited an upward curvature at higher surface densities. We constructed and analyzed the internal pressure profile of one GMC, K001A, that best exemplified this profile type. We demonstrated this cloud to be in a state of virial equilibrium throughout its entire interior structure, supporting the hypothesis that this GMC is likely in pressurized hydrostatic equilibrium across all cloud boundaries, including its outermost one.  

The basic similarity of the K001A' s pressure profile to those of the individual GMCs shown in Figure \ref{PressureDiagramM31_individualGMCs} further suggests the possibility  that the structure of K001A may be representative of the internal structures of many GMCs in Figure \ref{PressureDiagramM31_individualGMCs}. 
Inspection of the figure shows that numerous GMC pressure relations, like that of K001A, lie on or very near the virial line. The profiles of most GMCs, however, lie between the virial and  the bound/unbound boundary lines indicating that these clouds are bound. 
Since this figure is somewhat crowded with multiple overlapping profiles, we separately plot the profiles for two other GMCs, K098A and K213A, in Figure \ref{PvsSD2GMCs} to more clearly illustrate how comparable the natures of GMC pressure profiles in M31 can be. 
The profile of K213A closely resembles that of K001A in Figure \ref{PvsSDK001} with both clouds showing the low surface density points to be coincident the virial line, within the relatively small internal uncertainties. At higher surface densities K213A deviates upward from the virial relation, like K001A, similarly suggesting an internal structure described by pressurized virial equilibrium across all cloud boundaries or radii. We also plot the pressure profile for the GMC, K098A, a cloud whose profile more typically lies above the virial line. The pressure profile of K098A  closely matches the shape of K213A's profile where they overlap in surface density. Moreover, when accounting for the mutual internal errors (shown in the figure), the profiles of K098A and K213A are essentially coincident within the formal uncertainties. This is consistent with and supports the hypothesis that these GMCs are in virialized pressure and hydrostatic equilibrium across their internal structure.

A similar conclusion may apply to most of the sources in Figure \ref{PressureDiagramM31_individualGMCs},
although there is some variation in pressure profile shapes across the GMC population. For example, 
there are some sources whose profiles have less pronounced or no upward curvature as can be ascertained in the figure and there are a handful of sources with profiles that decline at the higher surface densities. Nonetheless, the profile shapes of sources K001A,  K098A and K213A are typical of the bulk of the GMCs, independent of their exact position on the diagram. Moreover, the similarity of the profiles may extend beyond their shapes to their true locations on the diagram when one considers that the positions of the individual GMCs in Figure \ref{PressureDiagramM31_individualGMCs}  are also 
subject to additional uncertainties due to systematic effects. The dominant one being the uncertainty in the CO mass calibration, $\alpha$(CO), which for the M31 clouds is on the order of 40\% (Viaene et al. 2021). Furthermore, the locations of the virial and  bound/unbound boundary lines themselves are also uncertain due to the assumptions of spherical geometry and a specific density structure used to calculate their positions. However, the systematic uncertainties enumerated above are not important for the assessments of the shapes of the GMC pressure profiles studied here. The relevant uncertainties for that purpose are the internal uncertainties that are tied to the image noise in the observations that was discussed earlier. 
The fundamental property of the GMCs examined here is that their pressure profiles increase in an approximately power-law fashion with surface density. That the profiles are also generally parallel to the virial and bound/unbound boundary lines indicates that $p_{int} \sim \Sigma^2$ for these GMCs and therefore that they are likely in approximate  virial balance at all depths. For most GMCs whose pressure profiles exhibit upward curvature at the highest surface densities, the upward trajectories are also consistent with the expectations of virial or hydrostatic balance with pressure confinement provided in the innermost regions by the weight of their outer layers.

 Additional support for this interpretation of the internal dynamical nature of these  M31 GMCs comes from \thco observations of the clouds. This is illustrated by the two examples  in Figure \ref{PvsSD_12+13_2GMCs} where we present both the \twco and \thco pressure profiles for the GMCs, K001A and K270A. Here the \thco pressures were calculated for the same areas as the \twco and the pressure differences are due to the difference in velocity dispersions between the two isotopologues.  We also carried out a separate analysis where the \thco data was independently extracted using the same methodology as employed for the \twco data analysis.
Out of the 48 GMCs in our sample, 31 were found to have \thco emission that was both sufficiently spatially resolved and of sufficiently large dynamic range in column density  to construct 5-point pressure profiles, similar to those created from the \twco data. Overall the 31 \thco pressure profiles closely resemble and align with the \twco profiles, similar to the two examples illustrated in Figure \ref{PvsSD_12+13_2GMCs}.  The similarity between the profiles from the \twco and \thco extractions suggests that the same underlying physics driving the \twco relations—namely, hydrostatic/virial equilibrium at all cloud depths—is likely responsible for the overall structure of the \thco pressure relations as well.

\begin{figure}[]
\centering
\includegraphics[width=0.5\hsize]{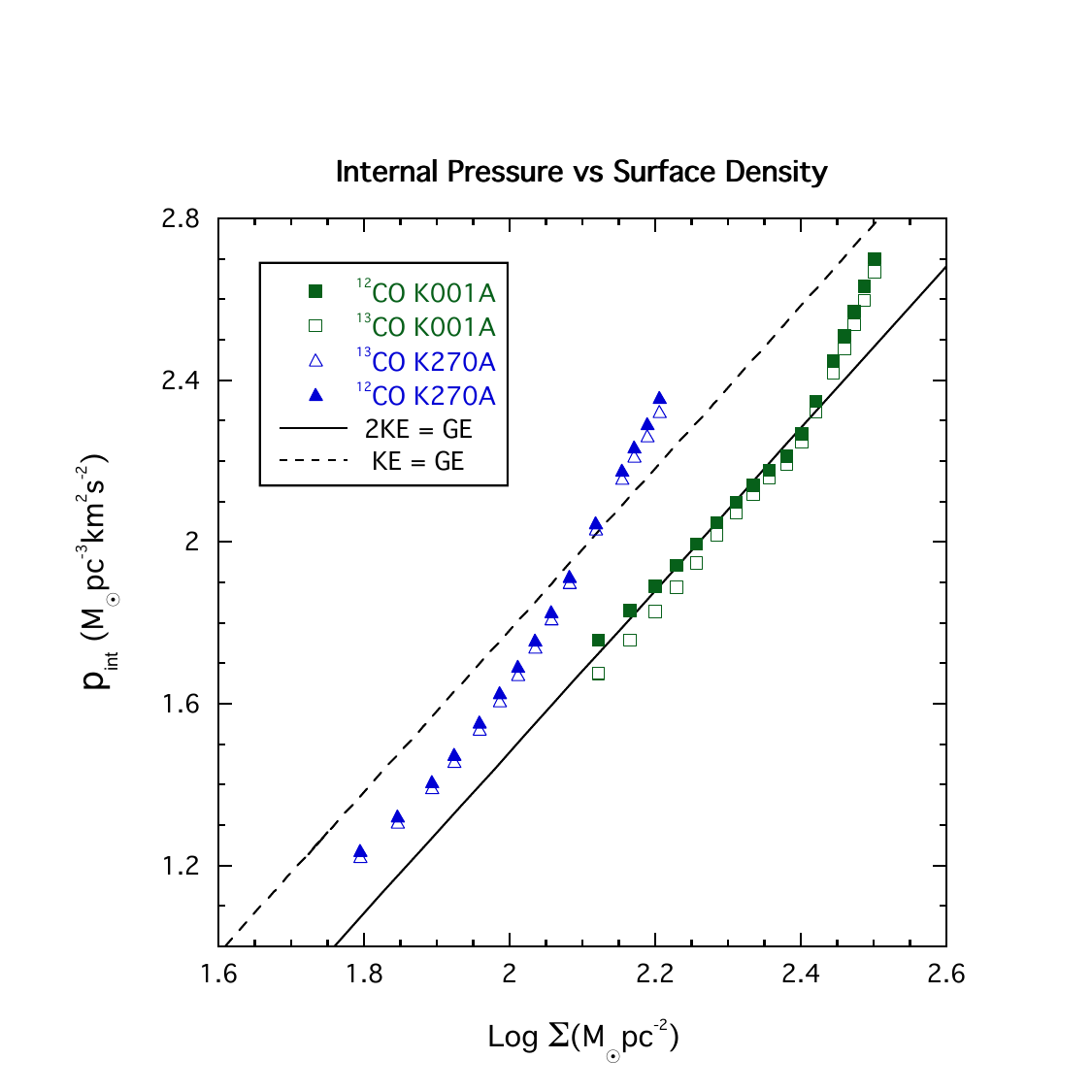}
\caption{Internal pressure--surface density($p_{int}$--$\Sigma$)  relations for \twco (solid symbols) and \thco (open symbols)  for the GMCs K001A (squares) and K270A  (circles) in M31. The  \twco and \thco relations for both sources are very similar in shape and position and coincident within the internal uncertainties.   Otherwise, same as  figure \ref{PvsSDK001}. \label{PvsSD_12+13_2GMCs}} 
\end{figure}

As noted earlier in this paper, 57\% of the GMCs in M31 appear unbound, with their kinetic energies exceeding their gravitational binding energies. In our analysis, we were able to evaluate pressure profiles for only 11 of these unbound clouds, or 16\% of the total. We find them to be of similar nature to the rest of the GMC pressure profiles. Consider that our analysis of the virial diagram for both bound and unbound GMCs in M31 revealed that, for the clouds that could be traced to their outermost physical boundaries, the pressure required for their confinement was comparable to the expected mid-plane pressure of  M31. Based on these findings, we conclude that, while most GMCs in M31 (by number) may have bulk kinetic energies greater than their gravitational binding energies, it is quite plausible that they are all confined by the pressure of the surrounding interstellar medium in the M31 disk  and are not freely expanding.  Some of these GMCs may also be in a state of virial, if not hydrostatic, equilibrium.

There is an important caveat regarding the validity of our pressure analysis. Our underlying premise is that the CO velocity dispersions are due to relatively small scale (l $< $10 pc) turbulent motions. If the dispersions were instead due to bulk systematic motions such as global collapse (e.g., Goldreich \& Kwan 1974; Zamora-Aviles et al. 2024) then our interpretation of the virial and pressure diagrams would not necessarily be unique. For example, globally collapsing clouds can produce profiles that parallel the virial relation on the virial and pressure plots used here (Shu 1977, Krumholz et al. 2025). However, such clouds would not be expected to simultaneously produce profiles with upward curvature at high surface densities (Krumholz et al. 2025). 
For those GMCs with cloud profiles absent of such upward curvature, their dynamical nature is somewhat  ambiguous and would require additional information for a more definitive assessment of their dynamical state. 


\subsection{Broader Implications}

 The Milky Way and Andromeda galaxies exhibit several key differences in their global physical properties. The metallicity gradient of the Milky Way is much steeper than that of M31 (Bosomworth et al. 2025).  The Milky Way's disk  has better defined spiral structure, while M31's disk is dominated by a prominent ring of active star formation. The scale height of M31's disk is significantly larger than that of the Milky Way (Dalcanton et al. 2023). These differences may all be attributable to the fact that during the last 2-3 billion years  M31 has experienced collisions with its companion galaxy M32, (Block et al. 2006; Dierickx et al. 2014, D'Souza \& Bell 2018) while during the same period the Milky Way has evolved in a more secular manner. 
 
 Despite these differences, studies of the molecular clouds in both galaxies have shown notable similarities  in their physical properties (e.g. Vogel et al. 1987, Lada et al. 1988, Loinard et al. 1999).  Recently, in the most comprehensive study of GMCs in M31 to date, many of the fundamental physical properties of its GMCs  were found to be "almost indistinguishable"  from those of local Milky Way clouds (Lada et al. 2024). The results of the present study, when combined with those obtained recently for Milky Way clouds by Keto (2024), further support the concept of closely similar GMC populations in the two galaxies. In particular, the individual GMCs in both galaxies show evidence for hydrostatic equilibrium characterized by gravitational binding within a confining internal pressure that is provided by the weight of the GMC itself. There is also an overall pressure balance across the outer boundaries of the clouds where the mid-plane pressure of each galaxy's disk provides the confining pressure. 

The broader implications of these findings for the  evolution of molecular clouds are not presently clear. 
Although many clouds in M31 appear to be in hydrostatic equilibrium, they remain turbulent, dynamic systems likely undergoing changes over time. Hydrostatic equilibrium does not guarantee long-term stability. For example, a cloud may be in an unstable equilibrium state, similar to a critically stable Bonner-Ebert sphere.  Once disturbed, the cloud would collapse or disperse on a free-fall timescale. However, this scenario is unlikely for the M31 GMCs studied here. If their equilibrium lasted less than a free-fall time and we are observing them at random stages of evolution, it would suggest that  most of these GMCs just happened to synchronize themselves to be in this short-lived phase at the exact moment we observed them. This seems unlikely and instead suggests that the clouds are in a quasi-stable state lasting on the order of, or longer than, a free-fall or dynamical time. 

The apparent  stability suggested here could help ease the long recognized problem that Milky Way molecular clouds are observed to form stars at a significantly lower rate than predicted from simple theoretical considerations for clouds evolving on a free-fall timescale  which is on the order of 3 Myrs for typical GMC number densities of 100 cm$^{-3}$ (e.g., Zuckerman \& Palmer 1974;  Zuckerman \& Evans 1974; Krumholz \& McKee 2005).  With a molecular mass one quarter that of the Milky Way (Dame et al. 1993) and lower global star formation rate this issue is also problematic for M31. Clouds in approximate hydrostatic equilibrium would be expected to evolve on somewhat longer timescales, consistent with modern estimates of cloud lifetimes in galaxies of up to 30 Myr (e.g., Schinnerer \& Leroy 2024 and references therein). 

Feedback from star formation is believed to be highly effective at disrupting clouds on timescales of ~5 - 10 Myr after the onset of star formation (e.g., Leisawitz et al. 1979). As a result, one might expect to observe at least some GMCs that are unbound and breaking apart. However, would we even recognize clouds in such a state? A recent study of the Sco-Cen star forming region in the Milky Way suggests that a significant fraction of the original molecular gas survives the disruption process. This surviving gas, however, is more dispersed and fragmented into smaller units that lack the fraction of high-density material typical of their progenitor GMC (Alves et al. 2025). Surviving clouds smaller than about 10 pc in size would be filtered out in the present study due to our selection criteria. Yet the lifetime of any surviving molecular gas we could detect would be perceived to be longer than a free-fall time. In this context, it is interesting to note that Lada et al. (2024) identified two distinct physical types of GMCs in Andromeda, which they classified as "dense" and "diffuse" GMCs. This classification was based on the presence or absence of significant \thco emission, respectively. 
Although diffuse GMCs make up 62\% of the total GMC population, they account for only 27\% of the total mass. The average sizes of diffuse GMCs are roughly a factor of 2 less than those of the dense GMCs. Furthermore, only 25\% of diffuse GMCs appear to be gravitationally bound in the absence of external pressure. Of the 48 GMCs for which we obtained reliable pressure profiles, only 11 were diffuse and of these 7 were bound objects. The average surface densities of diffuse GMCs are nearly a factor of 2 lower than those of dense GMCs, but individually they do not typically span a sufficiently broad range of internal surface densities to appear on the pressure plot in figure \ref{PressureDiagramM31_individualGMCs}. Nonetheless, as argued earlier, these clouds too may be  quasi-stable, pressure-confined objects. Their smaller sizes and densities may also indicate that some of these diffuse GMCs are remnants of disruptive feedback events similar to the clouds associated with the Sco-Cen OB association in the Milky Way. But it is also possible that they represent the initial stages of molecular cloud evolution prior to active star formation, corresponding to the empirical evolutionary stages I and II  described by Fukui (1999). In this case their lifetimes (6-20 Myr; Kawamura et al. 2009) would also be longer than the free-fall timescale and thus consistent with their relative dynamical stability inferred here. However, a more definitive statement about the evolutionary nature of the diffuse GMCs would be speculative at this time. 

Beyond Andromeda and the Milky Way our results are also consistent with observations of nearby galaxies.  For example, in a study of the galaxy NGC 300, Faesi et al. (2018) argued that differences between the GMC surface densities of NGC 300 and the galaxy M 51 could be accounted for by the differences in the estimated mid-plane pressures in the two galaxies if the GMCs in both galaxies were confined by these ambient disk pressures. In particular, our findings support the study by Sun et al. (2020) which investigated the dynamical states of molecular clouds in 28 nearby, but more distant, star forming galaxies. On spatial scales of 1 kpc$^2$ their study found that the turbulent pressure of molecular gas was correlated with, but often exceeded, estimates of the mid-plane pressures in these galaxies. This  suggested that the molecular gas was overpressurized relative to its surrounding environment. However, Sun et al. surmised that their mid-plane pressure estimates, which scaled with the square of the total (atomic + molecular) gas surface density, were likely underestimating the actual pressures felt by the molecular gas on $\sim$ 100 pc scales. They hypothesized that this was due to the clumpiness of the molecular gas on 1 kpc$^2$  scales which diluted its contribution to the mid-plane pressure estimates. After accounting  for this effect Sun et al. found that, on cloud scales, the ambient mid-plane pressure agreed well with the turbulent pressures of the molecular gas. Our findings showing that M31 GMCs are in rough pressure equilibrium with the ambient disk pressure corroborates this analysis.  This is in large part because our estimates of mid-plane pressure in M31 are more representative of the pressure at the individual cloud surfaces since they were computed for smaller spatial scales (0.5 kpc²) in a galaxy where atomic hydrogen dominates the total gas surface density in the disk.

\section{Summary}

 We analyzed \twco and \thco observations from a deep SMA survey of the Andromeda galaxy to examine the role of pressure, both external and internal, in determining the structure, stability and dynamical nature of the galaxy's GMC population. From an analysis of global properties of the GMCs we confirmed earlier results that suggested that the kinetic energies  exceed the gravitational binding energies for a large fraction, if not most, GMCs. Further analysis indicated that the range of external pressure that would be required to confine the GMC population was too large to be provided by the mid-plane pressure in the disk of a galaxy such as M31. The exception being where the nominal image noise was sufficiently low to enable the clouds to be traced to  depths (A$_{\rm v} \approx$ 1.0 mag) near their outermost molecular boundaries. For these clouds the external pressures required for confinement are comparable to the estimated mid-plane pressure of M31.  Furthermore, for the remaining GMCs  the required confining pressures were found to systematically increase with the surface densities of the empirically adopted, noise fixed, outer cloud boundaries. This suggests that these clouds are likely pressure confined by the weight of their (undetected) outer envelopes. Together these considerations indicate that the GMCs in M31 are likely in or close to virial equilibrium across all radii.
 
To test this hypothesis, we developed a methodology that allowed us to use CO observations to measure internal cloud pressures as a function of cloud scale or depth in 48 of our best-resolved GMCs. We found that the pressure profiles derived from \twco and \thco emission were well behaved and strikingly similar, with internal pressure gradients increasing steeply in a power-law fashion with  cloud surface density  (i.e., $ p_{int} \sim \Sigma^{2}$) and, in numerous cases, with additional upward curvature at the highest surface densities.  We argue that these behaviors are in agreement with the theoretical expectations of hydrostatic equilibrium and supports the hypothesis that these GMCs in M31 may be in or near virial equilibrium across their entire structure, including the outermost boundaries.
 

 \section{Acknowledgements}

  We acknowledge that the data analyzed here were obtained with the Submillimeter Array which is a joint project between the Smithsonian Astrophysical Observatory and the Academia Sinica Institute of Astronomy and Astrophysics and is funded by the Smithsonian Institution and the Academia Sinica. We thank Professor Pauline Barmby for providing us with the Spitzer measurements used to derive the stellar surface density across M31. MRK acknowledges support from the Australian Research Council through Laureate Fellowship award FL220100020.

\end{document}